\documentstyle[preprint,aps]{revtex}   
\begin{document}
\title{Phenomenology of the Two Higgs Doublet
Sector of a Quark-Lepton Symmetric Model}
\draft
\preprint{UM-P-94/21; OZ-94/8}
\author{C. C. Lassig\footnote{ccl@tauon.ph.unimelb.edu.au}
and R. R. Volkas\footnote{U6409503@hermes.ucs.unimelb.edu.au}}
\address{Research Centre for High Energy Physics, \\
School of Physics, The University of Melbourne, \\ 
Parkville, Victoria 3052. Australia}
\date{\today}
\maketitle
\begin{abstract}
In the simplest examples of models with a
discrete quark-lepton symmetry, an electroweak symmetry
breaking sector with more than one Higgs doublet is necessary to 
obtain the correct mass relations between quarks and leptons. A two Higgs
doublet model version has flavour-nonconserving Yukawa couplings,
which are proportional to the masses of the quark-lepton symmetric partners
of the fermions. We describe how flavour changing
leptonic decays can occur, with
branching ratios not far beyond that currently measurable, enabling
investigation of the phenomenology of such models.
\end{abstract}
\pacs{}

\section{Introduction}
The Quark-Lepton Symmetric extension of the Standard Model involves
the addition of leptonic colour to the symmetry group,
in order that the lagrangian can be made to exhibit a discrete
symmetry between quarks and leptons \cite{FootLew}. The symmetry group then
becomes $G_{q\ell} \equiv $
SU(3)$_{\ell}\times$SU(3)$_q\times$SU(2)$_L\times$U(1)$_X$.
The leptonic colour group SU(3)$_{\ell}$ can be spontaneously broken at a scale
as low as about a few TeV. 
In the simplest models, SU(3)$_{\ell}$ breaks down to SU(2)$'$, which remains
unbroken, and acts to bind the exotic colour partners of 
leptons into hadron-like states.
Then for the breaking of the SU(2)$_L$ 
electroweak symmetry, the minimal choice of a single Higgs doublet
as for the Standard Model gives rise to unsuitable relations 
between the masses of the quarks and leptons \cite{LeVolk}. It is thus necessary
to use a slightly expanded Higgs sector, and this paper will 
examine some of the phenomenological consequences of choosing the
next simplest possibility, that of two Higgs doublets.
The present paper extends the analysis given in Ref.\cite{LeVolk}.
It is important to note however, that these unsatisfactory mass relations
can be avoided by having a different form of quark-lepton symmetric model,
such as one in which leptonic colour is broken completely, or by incorporating
a left-right symmetry (see papers 9 and 13 in Ref.\cite{FootLew}).

Two Higgs doublet models (2HDMs) are usually constructed so that tree-level
flavour-changing neutral processes are absent \cite{WeinGlash}, although
a generic 2HDM would feature these processes. In the two Higgs doublet
version of quark-lepton symmetric models, tree-level flavour changing
processes are, by contrast, unavoidable.
As the lagrangian is symmetric between quarks and leptons, the flavour-changing
leptonic terms are dependent on quark masses, and vice-versa, so 
leptonic processes are of more interest as the heavy quark masses
give rise to proportionally large branching ratios.

There are in fact a number of alternative ways of incorporating
two Higgs doublets into quark-lepton symmetric models,
but here we have concentrated
on the model which gives the larger contribution to flavour-nonconserving
decays, the Model 1 of Ref.\cite{LeVolk}, the main results concerning which
will be briefly stated in Sec.II of this paper.
Other models, such as the Model 2 of Ref.\cite{LeVolk},
lead to leptonic Yukawa couplings with a dependence on lepton masses as well
as quark masses. As stated above, it is the couplings proportional to quark
masses that are of interest, as these lead to flavour-changing reactions.

Thus the main focus of this paper will be on flavour-changing leptonic
decays, in particular those of muons since these give the most  
opportunity for experimental investigation. First in Sec.III the neutral Higgs
mediated decay $\mu^- \rightarrow e^- e^+ e^-$ will be
investigated at tree-level, as will the charged-Higgs contribution
to the decay $\mu^- \rightarrow \nu_{\mu} e^- \bar{\nu_e}$.
In Sec.IV  a one-loop calculation of the process $\mu \rightarrow
e \gamma$ will be performed. This will lead on to an examination of the
effect on the muon anomalous magnetic moment. Finally, the effect
of the quark-lepton symmetric 2HDM contribution to $b \rightarrow
s \gamma$ will briefly be considered.

\section{Yukawa Couplings in a Quark-Lepton Symmetric Model}
The symmetry-breaking sector in the two Higgs doublet 
version of the quark-lepton symmetric model we will consider here
consists of Higgs fields with the following
quantum numbers under the symmetry group $G_{q\ell}$ \cite{LeVolk}:
\begin{equation}
\chi_1 \sim (1,3,1)(-2/3),\; \chi_2 \sim (3,1,1)(2/3),\; 
\phi_1 \sim (1,1,2)(1),\; \phi_2 \sim (1,1,2)(-1),
\end{equation}
where $\chi_1 \leftrightarrow \chi_2$ and $\phi_1 \leftrightarrow \phi_2$
under the discrete quark-lepton symmetry.

Symmetry breaking occurs in two stages.
In the first stage, $\chi_2$ acquires a 
non-zero vacuum expectation value (VEV), which breaks the leptonic colour
group SU(3)$_{\ell}$ down to SU(2)$'$, and thus also
breaks the discrete symmetry.
The many other interesting features of quark-lepton symmetric models, such
as the exotic leptons of the unbroken SU(2)$'$ group, and the new
gauge bosons which appear, are dealt with elsewhere \cite{FootLew}.
To achieve electroweak symmetry breaking, both $\phi_1$ and $\phi_2$
acquire VEVs. The corresponding Yukawa Lagrangian is 
\begin{eqnarray}
{\cal L}_{\rm Yuk} & = & \lambda_1 [\overline{(Q_L)^c}Q_L \chi_1 +
\overline{(F_L)^c}F_L \chi_2]+\lambda_2 [\overline{(u_R)^c}d_R \chi_1 +
\overline{(e_R)^c}\nu_R \chi_2] \nonumber\\
& & + \Lambda_1 [\overline{Q_L}d_R \phi_1 + \overline{F_L}\nu_R \phi_2]
+ \Lambda'_1 [\overline{Q_L}d_R \phi_2^c + \overline{F_L}\nu_R \phi_1^c]
\nonumber\\
& & + \Lambda_2 [\overline{Q_L}u_R \phi_1^c + \overline{F_L}e_R \phi_2^c]
+ \Lambda'_2 [\overline{Q_L}u_R \phi_2 + \overline{F_L}e_R \phi_1] + {\rm H.c.},
\label{TotYuk}
\end{eqnarray}
where $Q_L$ and $F_L$ represent the left-handed quark and lepton weak doublets
respectively, and $\phi^c$ is the charge-conjugate of the scalar field,
$\phi^c = i \tau_2 \phi^*$.

Electroweak symmetry breaking gives rise to physical fields from the 
two electroweak doublets as in normal 2HDMs \cite{Hunter}.
Thus there is a charged
field $H^{\pm}$, a CP-odd neutral field $\eta$, and two CP-even
neutral fields, $h_1$ and $h_2$.
There is also one neutral physical Higgs field: that part of $\chi_2$ which
survives after quark-lepton symmetry breaking. Following Ref.\cite{LeVolk},
we will neglect mixing between this field and $\eta$,
$h_1$ and $h_2$, since $\chi_2$ is expected to be much more massive than
any of these electroweak fields.

The resulting Yukawa interactions for these physical fields are then
\cite{LeVolk}
\begin{eqnarray}
& {\cal L}_{\rm Yuk}^+ = & \frac{1}{u} \bar{d_L}m_e u_R H^-
+ \frac{1}{u} \bar{u_L} m_{\nu}^{\rm Dirac} d_R H^+
+ \frac{1}{u} \bar{\nu_L} m_u e_R H^+
+ \frac{1}{u} \bar{e_L} m_d \nu_R H^- + {\rm H.c.}, \\
& {\cal L}_{\rm Yuk}^{\eta} = & \frac{i}{\sqrt{2}u} \bar{u_L}
m_e u_R \eta + \frac{i}{\sqrt{2}u} \bar{e_L} m_u e_R \eta +
\frac{i}{\sqrt{2}u} \bar{d_L} m_{\nu}^{\rm Dirac} d_R \eta +
\frac{i}{\sqrt{2}u} \bar{\nu_L} m_d \nu_R \eta + {\rm H.c.}, \\
& {\cal L}_{\rm Yuk}^h = & \frac{1}{\sqrt{2}u} \bar{u_L}
\Big[ m_u \{ \cos(\omega-\varphi)h_1 + \sin(\omega-\varphi)h_2\} \nonumber\\
&\ &\qquad\quad + m_e \{ - \sin(\omega-\varphi)h_1 + \cos(\omega-\varphi)h_2\}
\Big] u_R \nonumber\\
& & + \frac{1}{\sqrt{2}u} \bar{e_L} \Big[ m_u \{\sin(\omega-\varphi)h_1
-cos(\omega-\varphi)h_2\} \nonumber\\
&\ &\qquad\quad + m_e \{\cos(\omega-\varphi)h_1 + \sin(\omega-\varphi)h_2\}
\Big] e_R \nonumber\\
& &  +\frac{1}{\sqrt{2}u} \bar{d_L} \Big[ m_d \{cos(\omega-\varphi)h_1
+ \sin(\omega-\varphi)h_2\} \nonumber\\
&\ &\qquad\quad + m_{\nu}^{\rm Dirac} \{\sin(\omega-\varphi)h_1 -
\cos(\omega-\varphi)h_2\} \Big] d_R \\
& & + \frac{1}{\sqrt{2}u} \bar{\nu_L} \Big[ m_d \{-\sin(\omega
-\varphi)h_1 + \cos(\omega-\varphi)h_2\} \nonumber\\
&\ &\qquad\quad + m_{\nu}^{\rm Dirac} \{\cos(\omega-\varphi)
h_1 + \sin(\omega-\varphi)h_2\} \Big] \nu_R + {\rm H.c.} \nonumber
\end{eqnarray}
where $u \equiv \sqrt{u_1^2 + u_2^2}$, with $u_1$ and $u_2$ being the VEV's of
$\phi_1$ and $\phi_2$, respectively. Also in the above, $\tan \omega \equiv
u_2 / u_1$, and $\varphi$ is a mixing angle relating the mass-eigenstate 
fields $h_1$ and $h_2$ to the CP-even parts of $\phi_1$ and $\phi_2$.

These Yukawa lagrangians are written in terms of the gauge eigenstate fermion
fields, eg. $u$ for the charge 2/3 quarks.
To obtain the corresponding mass-eigenstate fields, $U$, the unitary 
diagonalisation matrices $V_{L,R}^u$ are introduced:
\begin{equation} 
U_{L,R} \equiv V_{L,R}^u u_{L,R}
\end{equation}
and the diagonal mass matrix $M_u = V_L^u m_u V_R^{u\dagger}$.
Analogous relations of course apply to the other kinds of fermions, $d$, $e$
and $\nu$.

Writing the Yukawa lagrangians in terms of the mass-eigenstate fields
[and using the notation $V_{L,R}^{fg} \equiv V_{L,R}^f V_{L,R}^{g\dagger}$,
so that for instance the Cabibbo-Kobayashi-Maskawa (CKM) quark mixing matrix
is $V_L^{ud}$], we obtain for the flavour-changing interactions
\begin{eqnarray}
& {\cal L}_{\rm Yuk}^+ = & \frac{1}{u} \bar{U} \Big( V_L^{u\nu}M_{\nu}
V_R^{d\nu\dagger}\gamma_+ + V_R^{ue}M_e V_L^{de\dagger}\gamma_-   
\Big) D H^+ \nonumber\\
& & + \frac{1}{u} \bar{N} \Big( V_L^{u\nu\dagger}M_u V_R^{ue}\gamma_+
+ V_R^{d\nu\dagger}M_d V_L^{de}\gamma_- \Big) E H^+ + {\rm H.c.},\\
\label{HYuk}
& {\cal L}_{\rm Yuk}^\eta = & \frac{i}{\sqrt{2}u} \bar{U} \Big( 
V_L^{ue}M_e V_R^{ue\dagger}\gamma_+ \Big) U \eta + \frac{i}{\sqrt{2}u}
\bar{E} \Big( V_L^{ue\dagger}M_u V_R^{ue}\gamma_+ \Big) E \eta \nonumber\\
& & + \frac{i}{\sqrt{2}u} \bar{D} \Big(V_L^{d\nu}M_{\nu}V_R^{d\nu\dagger}
\gamma_+ \Big) D \eta + \frac{i}{\sqrt{2}u} \bar{N} \Big( V_L^{d\nu\dagger}
M_d V_R^{d\nu}\gamma_+ \Big) N \eta + {\rm H.c.},\\
\label{nYuk}
& {\cal L}_{\rm Yuk}^h = & \frac{1}{\sqrt{2}u} \bar{U} \Big( V_L^{ue}
M_e V_R^{ue\dagger}\gamma_+ \Big) U \Big[ -\sin(\omega-\varphi)h_1
+\cos(\omega-\varphi)h_2 \Big] \nonumber\\
& & + \frac{1}{\sqrt{2}u} \bar{E} \Big( V_L^{ue\dagger}M_u V_R^{ue}\gamma_+
\Big) E \Big[ \sin(\omega-\varphi)h_1-\cos(\omega-\varphi)h_2 \Big] \nonumber\\
& & + \frac{1}{\sqrt{2}u} \bar{D} \Big( V_L^{d\nu}M_{\nu}V_R^{d\nu\dagger}
\gamma_+ \Big) D \Big[ \sin(\omega-\varphi)h_1-\cos(\omega-\varphi)h_2 \Big] \\
\label{hYuk}
& & + \frac{1}{\sqrt{2}u} \bar{N} \Big( V_L^{d\nu\dagger}M_d V_R^{d\nu}
\gamma_+ \Big) N \Big[ -\sin(\omega-\varphi)h_1+\cos(\omega-\varphi)h_2 \Big]
+ {\rm H.c.} \nonumber
\end{eqnarray}

It is important to observe that down-quark sector neutral flavour-changing
vertices are proportional to neutrino Dirac masses. Because of the severe
experimental upper bounds on neutrino masses, and because they are Dirac
particles in our model, down-quark sector effects are negligible. This is
pertinent because the constraint from $K^0 - \bar{K^0}$ mixing would 
otherwise be very stringent (see Ref.\cite{LeVolk} for further discussion).
We will thus concentrate on processes in the charged-lepton and up-quark
sectors.

In the rest of this paper, the following notation will be used:
\begin{eqnarray}
(u_1,u_2,u_3) \equiv & (u,c,t) ;\; (d_1,d_2,d_3) \equiv & (d,s,b);\nonumber\\
(e_1,e_2,e_3) \equiv & (e,\mu,\tau) ;\; (\nu_1,\nu_2,\nu_3) \equiv
& (\nu_e,\nu_{\mu},\nu_{\tau});\nonumber
\end{eqnarray}
\begin{equation}
y_f \equiv \frac{m_f^2}{m_H^2} ; \; z_f \equiv \frac{m_f^2}{m_{\eta}^2} ;\;
w_f^{(1,2)} \equiv \frac{m_f^2}{m_{h_{1,2}}^2};\nonumber
\end{equation}
\begin{equation}
\frac{1}{m_{\pm}^2} \equiv \frac{\sin^2(\omega-\varphi)}{m_{h_1}^2}
+\frac{\cos^2(\omega-\varphi)}{m_{h_2}^2} \pm \frac{1}{m_{\eta}^2};\nonumber
\end{equation}
\begin{equation}
w_f^{\pm} \equiv \frac{m_f^2}{m_{\pm}^2}.\nonumber
\end{equation}

\section{Leptonic Decays at Tree Level}
\subsection{$\mu^- \rightarrow e^- e^+ e^-$}
At tree level, the flavour-changing decay $\mu^- \rightarrow e^- e^+ e^-$
can be mediated by the neutral Higgs particles, $\eta$, $h_1$ and $h_2$. The
branching ratio for this decay is (with the approximation 
$m_e^2 \ll m_{\mu}^2$),
\begin{eqnarray}
B(\mu \rightarrow e \bar{e} e) & \approx &
\frac{3}{4} \Big[ \frac{1}{4 m_-^4} \Big( \Big|
\sum_{k,l} m_{u_k} m_{u_l} V_{Lk1}^{ue*} V_{Rk2}^{ue} V_{Ll1}^{ue*}
V_{Rl1}^{ue} \Big|^2 + \Big| \sum_{k,l} m_{u_k} m_{u_l} V_{Rk1}^{ue*}
V_{Lk2}^{ue} V_{Rl1}^{ue*} V_{Ll1}^{ue} \Big|^2 \Big) \nonumber\\
& & + \frac{1}{6 m_+^4} \Big( \Big| \sum_{k,l} m_{u_k} m_{u_l} V_{Rk1}^{ue*}
V_{Lk2}^{ue} V_{Ll1}^{ue*} V_{Rl1}^{ue} \Big|^2 + \Big|
\sum_{k,l} m_{u_k} m_{u_l}
V_{Lk1}^{ue*} V_{Rk2}^{ue} V_{Rl1}^{ue*} V_{Ll1}^{ue} \Big|^2 \Big) \nonumber\\
& & + \frac{m_e}{m_{\mu}} \Big( \frac{1}{m_+^2m_-^2} \sum_{k,l,m,n}
m_{u_k} m_{u_l} m_{u_m} m_{u_n} (V_{Lk1}^{ue*} V_{Rk2}^{ue} V_{Ll1}^{ue*}
V_{Rl1}^{ue} V_{Rm1}^{ue*} V_{Lm2}^{ue} V_{Ln1}^{ue*} V_{Rn1}^{ue} \nonumber\\
&\ &\qquad\quad + V_{Rk1}^{ue*} V_{Lk2}^{ue} V_{Rl1}^{ue*} V_{Ll1}^{ue}
V_{Lm1}^{ue*} V_{Rm2}^{ue} V_{Rn1}^{ue*} V_{Ln1}^{ue}
+ {\rm c.c}) \Big) \nonumber\\
& & + \frac{1}{3} \frac{m_e}{m_{\mu}} \Big( \frac{1}{m_+^4} \sum_{k,l,m,n}
m_{u_k} m_{u_l} m_{u_m} m_{u_n} (V_{Rk1}^{ue*} V_{Lk2}^{ue} V_{Ll1}^{ue*}
V_{Rl1}^{ue} V_{Lm1}^{ue*} V_{Rm2}^{ue} V_{Rn1}^{ue*} V_{Ln1}^{ue} \nonumber\\
&\ &\qquad\quad + {\rm c.c})
\Big) \Big]
\label{BEEE0}
\end{eqnarray}

The above branching ratio involves many unknown parameters in the Higgs boson
masses and the mixing matrices. In order to get an indication of how large
$B(\mu \rightarrow e \bar{e} e)$ might be, we will suppose that all of the
mixing matrices display a hierarchical structure similar to that of the CKM
matrix, which can be written in the qualitative form
\begin{equation}
V \sim \left( \begin{array}{ccc}
		1	   & \epsilon	& \epsilon^3 \\
		\epsilon   & 1		& \epsilon^2 \\
		\epsilon^3 & \epsilon^2 & 1
	      \end{array} \right),
\label{Vhier}
\end{equation}
where for the CKM matrix for instance, $\epsilon \sim 0.22$. The
branching ratio is then approximately
\begin{eqnarray}
B(\mu \rightarrow e \bar{e} e) & \sim & \frac{3}{4} \Big[ \frac{1}{4}
\Big| \frac{1}{m_-^2} m_c V_{L21}^{ue*} V_{R22}^{ue} (m_u V_{L11}^{ue*}
V_{R11}^{ue} + m_c V_{L21}^{ue*} V_{R21}^{ue} ) \Big|^2
+ \Big| L \leftrightarrow R \Big|^2 \nonumber\\
& & + \frac{1}{6} \Big| \frac{1}{m_+^2} m_c V_{R21}^{ue*} V_{L22}^{ue}
(m_u V_{L11}^{ue*} V_{R11}^{ue} + m_c V_{L21}^{ue*} V_{R21}^{ue}) \Big|^2
+ \Big| L \leftrightarrow R \Big|^2 \nonumber\\
& & + \frac{m_e}{m_{\mu}} \frac{1}{m_-^2 m_+^2} \Big( m_c^2 V_{L21}^{ue*}
V_{R22}^{ue} V_{R21}^{ue*} V_{L22}^{ue} (m_u V_{L11}^{ue*} V_{R11}^{ue}
+ m_c V_{L21}^{ue*} V_{R21}^{ue} )^2 + {\rm c.c.} \Big) \nonumber\\
& & + \Big(L \leftrightarrow R \Big) \Big] 
\label{BEEE1}
\end{eqnarray}
This branching ratio has the experimental upper bound 
$B(\mu \rightarrow e \bar{e} e) < 1.0 \times 10^{-12}$.

At this point, in order to get some idea of the relation of the branching
ratio to the Higgs boson masses and mixing matrix elements,
it is useful to assume
that the matrices are of the form Eq.~(\ref{Vhier}), and, for convenience, that
they share the same $\epsilon$ parameter. 
Of course, this loses some of the features of the unitarity of the mixing 
matrices, as well as the possibility of accidental cancellation, but
nevertheless at least a qualitative understanding should be attainable.
Eq.~(\ref{BEEE1}) can then be
written:
\begin{equation}
B(\mu \rightarrow e \bar{e} e) \sim \frac{3}{4}  m_c^2 \epsilon^2 
(m_u + m_c \epsilon^2)^2 \Big[ \frac{1}{2 m_-^4} + \frac{1}{3 m_+^4}
+ \frac{m_e}{m_{\mu}} \frac{4}{m_-^2 m_+^2} \Big]
\label{BEEE2}
\end{equation}
There are three extreme possibilities for the relations between the masses
of the various neutral scalars, each of which leads to a slightly different
relation between the mass and the $\epsilon$ parameter.

\noindent If $m_{\eta} \approx m_{h_1} \approx m_{h_2}$,
\begin{equation}
\frac{1}{m_-^2} \ll \frac{1}{m_+^2} \equiv \frac{2}{m_{\phi}^2}.
\label{phi}
\end{equation}
If $m_{\eta}^2 \ll m_{h_1}^2, m_{h_2}^2$,
\begin{equation}
\frac{1}{m_+^2} \approx -\frac{1}{m_-^2} \approx \frac{1}{m_{\eta}^2}.
\label{eta}
\end{equation}
If $m_{\eta}^2 \gg m_{h_1}^2, m_{h_2}^2$,
\begin{equation}
\frac{1}{m_-^2} \approx \frac{1}{m_+^2} \equiv \frac{1}{m_h^2}.
\label{aitch}
\end{equation}
The dependence of the branching ratio on $\epsilon$ for various choices of
the Higgs boson masses is shown for these three cases in Fig.~(\ref{NeuBEEE}). 
The minimum value for the mass used, 48 GeV, is that obtained by the ALEPH
collaboration \cite{ALEPH}.
Only two lines for each $\epsilon$ value are visible, because the plots for
the conditions of Eq.~(\ref{eta}) and Eq.~(\ref{aitch}) are inseparable on
the scale used. It can be seen from this that an increase in the precision
of the measurement of the branching ratio by only a couple of orders of 
magnitude opens up for investigation vast new regions of the parameter space.
Indeed, a considerable range is already excluded. It can be seen that for
$\epsilon = 0.22$, corresponding to the CKM matrix, the Higgs boson mass
has to be greater than about 330 GeV.

Similar calculations can also be performed for the tauon as well, in such
processes as $\tau^- \rightarrow \mu^- \mu^+ \mu^-$ and 
$\tau^- \rightarrow e^- e^+ e^-$. 
Writing their branching ratios in the same form as Eq.~(\ref{BEEE2}), we
obtain
$\Big[ {\rm where} \; B_{\tau} \equiv B(\tau \rightarrow \nu_{\tau}e
\bar{\nu_e}) = 0.1793 \Big]$
\begin{eqnarray}
B(\tau \rightarrow \mu \bar{\mu} \mu) & \sim & B_{\tau}
\frac{3}{4} m_t^2 \epsilon^4 (m_c + m_t \epsilon^4)^2
\Big[ \frac{1}{2 m_-^4} + \frac{1}{3 m_+^4} + \frac{m_{\mu}}{m_{\tau}}
\frac{2}{m_+^2} \Big( \frac{2}{m_-^2} + \frac{1}{3 m_+^2} 
\Big) \Big],\\
B(\tau \rightarrow e \bar{e} e) & \sim & B_{\tau}
\frac{3}{4} m_t^2 \epsilon^{10} (m_c + m_t \epsilon^4)^2
\Big[ \frac{1}{2 m_-^4} + \frac{1}{3 m_+^4} + \frac{m_e}{m_{\tau}}
\frac{4}{m_-^2 m_+^2} \Big].
\end{eqnarray}
Using Higgs boson masses of the order
$\sim$ 100 GeV, and for the $\epsilon$ parameter
$\epsilon \sim 0.1$, these branching ratios are of the respective orders
$10^{-8}$ and $10^{-14}$
(here, as elsewhere in this paper, the value of $m_t = 150$ GeV is used
for the top quark mass, as this is in agreement with current experimental
data \cite{PDG}).
Experimentally, these branching ratios are less than
$1.7 \times 10^{-5}$ and $2.7 \times 10^{-5}$ respectively, therefore these
processes do not constrain our model to nearly the same extent as does
$\mu^- \rightarrow e^- e^+ e^-$.

\subsection{$\mu^- \rightarrow \nu_{\mu} e^- \bar{\nu_e}$}
At tree level in the Standard Model, the decay
$\mu^- \rightarrow \nu_{\mu} e^- \bar{\nu_e}$ is mediated by the $W$ boson.
In our model it can also be mediated by a charged scalar, $H^{\pm}$. This
can give a constraint on the various parameters (ie. mass and mixing matrices),
since the experimentally measured muon decay agrees so well with Standard
Model predictions. That is, the charged Higgs contribution must be of the
order of the uncertainty in the decay rate,
$\Gamma (\mu \rightarrow \nu_{\mu} e \bar{\nu_e}) = (2.99592 \pm 0.00005)
\times 10^{-16} {\rm MeV}$ \cite{PDG}.

Again with the approximation $m_e^2 \ll m_{\mu}^2$, and using hierarchical
mixing matrices as in Eq.~(\ref{Vhier}), the total decay rate, including 
both $W$ and Higgs contributions is
\begin{eqnarray}
\Gamma & = & \Gamma_W + \Gamma_H + \Gamma_{HW} \nonumber\\
& \sim & \Gamma_W \Big[ 1 + \frac{1}{4} \Big( y_c |V_{R21}^{ue}|^2
+ y_t |V_{R31}^{ue}|^2 \Big) \Big( y_c |V_{R22}^{ue}|^2 +
y_t |V_{R32}^{ue}|^2 \Big) \nonumber\\
& & + 2 \frac{m_e}{m_{\mu}} \frac{1}{m_H^2}
\Big( m_u V_{L11}^{ue} V_{L11}^{e\nu} V_{L11}^{u\nu*}
+ m_c V_{L21}^{ue} (V_{L11}^{e\nu} V_{L21}^{u\nu*} + V_{L12}^{e\nu}
V_{L22}^{u\nu*}) \nonumber\\
&\ &\qquad\quad  + m_t V_{L31}^{ue} (V_{L11}^{e\nu} V_{L31}^{u\nu*}
+ V_{L12}^{e\nu} V_{L32}^{u\nu*} + V_{L13}^{e\nu} V_{L33}^{u\nu*}) 
\Big) \nonumber\\
&\ &\qquad\quad \times \Big(m_c V_{R22}^{ue*} V_{L22}^{e\nu*} V_{L22}^{u\nu}
+ m_t V_{R32}^{ue*} (V_{L22}^{e\nu*} V_{L32}^{u\nu} +
V_{L23}^{e\nu*} V_{L33}^{u\nu}) \Big) \Big],
\label{HWH}
\end{eqnarray}
where
\begin{equation}
\Gamma_W = \frac{G_F^2 m_{\mu}^5}{192 \pi^2}.
\end{equation}
In Eq.~(\ref{HWH}), $V_L^{e\nu}$ is essentially the leptonic equivalent of
the CKM matrix, and must be used in the $W$ boson couplings.
 
Putting typical values for all the parameters ($m_H = 100$ GeV,
$\epsilon = 0.1$), the total contribution from the terms $\Gamma_H$ and
$\Gamma_{HW}$ is $(\Gamma_H + \Gamma_{HW})/\Gamma_W = 1.3 \times 10^{-8}$,
which is over three orders of magnitude less than
the experimental uncertainty, $\Delta \Gamma / \Gamma_W \approx
1.8 \times 10^{-5}$.

At this point it should be noted that the minimum mass for the charged Higgs
boson used in this paper is 45.3 GeV, as obtained by the ALEPH collaboration
\cite{ALEPH} with the assumption that the branching ratio $B(H^+ \rightarrow 
\tau^+ \nu) = 1$. In this particular 2HDM, with its quark-lepton symmetry,
leptonic decays are proportional to the square of quark masses (in particular
the top quark mass for $\tau^+ \nu$), whereas hadronic decays are proportional
to the squares of lepton masses. Thus here we can take $B(H^+ \rightarrow
\tau^+ \nu) = 1$, and so use the limit $m_H > 45.3$ GeV.

\section{One-Loop Processes}
\subsection{$\mu \rightarrow e \gamma$}
One-loop processes such as $\mu \rightarrow e \gamma$ and $b \rightarrow
s \gamma$ have been extensively studied, both in terms of the Standard Model
\cite{ILim} and 2HDMs \cite{bsgamma}.
In the two Higgs doublet quark-lepton symmetric model, the
branching ratio for the decay $\mu \rightarrow e \gamma$ is approximately
[from Eq.~(\ref{BEG0}) of the Appendix,
using the fact that neutrino masses are very small,
and the approximation $m_e^2 \ll m_{\mu}^2$]:
\begin{eqnarray}
B(\mu \rightarrow e \gamma) & \approx & \frac{\alpha}{96 \pi}
\Big[ \Big| \sum_k \Big( y_{d_k} V_{Lk1}^{de*} V_{Lk2}^{de} +
\frac{m_e}{m_{\mu}} y_{u_k} V_{Rk1}^{ue*} V_{Rk2}^{ue} -
w_{u_k}^+ V_{Lk1}^{ue*} V_{Lk2}^{ue} \nonumber\\
& & - \frac{m_e}{m_{\mu}} w_{u_k}^+ V_{Rk1}^{ue*} V_{Rk2}^{ue} +
9 \sum_{i,l} \frac{m_{e_i}}{m_{\mu}} \sqrt{w_{u_k}^- w_{u_l}^-}
V_{Lk1}^{ue*} V_{Rki}^{ue} V_{Lli}^{ue*} V_{Rl2}^{ue} \Big) \Big|^2 
\nonumber\\
& & + \Big| \sum_k \Big( y_{u_k} V_{Rk1}^{ue*} V_{Rk2}^{ue} +
\frac{m_e}{m_{\mu}} y_{d_k} V_{Lk1}^{de*} V_{Lk2}^{de} -
w_{u_k}^+ V_{Rk1}^{ue*} V_{Rk2}^{ue} \nonumber\\
& & - \frac{m_e}{m_{\mu}} w_{u_k}^+ V_{Lk1}^{ue*} V_{Lk2}^{ue} +
9 \sum_{i,l} \frac{m_{e_i}}{m_{\mu}} \sqrt{w_{u_k}^- w_{u_l}^-}
V_{Rk1}^{ue*} V_{Lki}^{ue} V_{Rli}^{ue*} V_{Ll2}^{ue} \Big) \Big|^2 \Big].
\label{BEG1}
\end{eqnarray}

Once again, the hierarchical structure Eq.~(\ref{Vhier}) can be used.
If the neutral scalars are all much heavier than the charged scalars, then
the branching ratio Eq.~(\ref{BEG1}) can be written
\begin{equation}
B(\mu \rightarrow e \gamma) \sim \frac{\alpha}{96 \pi} \frac{1}{m_H^4}
\Big| m_c^2 V_{R21}^{ue*} V_{R22}^{ue}
+ m_t^2 V_{R31}^{ue*} V_{R32}^{ue} \Big|^2
\label{BEGH}
\end{equation}
Alternatively, the neutral bosons could be much lighter --- then the same
possibilities as used for $B(\mu \rightarrow e \bar{e} e)$, Eqs.~(\ref{phi},
\ref{eta}, \ref{aitch}), 
can be used again:
\begin{eqnarray}
B(\mu \rightarrow e \gamma) & \sim & \frac{\alpha}{24 \pi}
\frac{1}{m_{\phi}^4} \Big[ \Big| m_c^2 V_{L21}^{ue*} V_{L22}^{ue}
+ m_t^2 V_{L31}^{ue*} V_{L32}^{ue} \Big|^2 + \Big| L \leftrightarrow
R \Big|^2 \Big] \\
B(\mu \rightarrow e \gamma) & \sim & \frac{\alpha}{96 \pi}
\frac{1}{m_{\eta}^4} \Big[ \Big| m_c^2 ( V_{L21}^{ue*} V_{L22}^{ue}
- 9 V_{L21}^{ue*} V_{R22}^{ue} V_{L22}^{ue*} V_{R22}^{ue} ) \nonumber\\
&\ &\qquad\quad + m_t^2 ( V_{L31}^{ue*} V_{L32}^{ue} - 9 
\frac{m_{\tau}}{m_{\mu}} V_{L31}^{ue*} V_{R33}^{ue} V_{L33}^{ue*}
V_{R32}^{ue} ) \Big|^2 + \Big| L \leftrightarrow R \Big|^2 \Big] \\
B(\mu \rightarrow e \gamma) & \sim & \frac{\alpha}{96 \pi}
\frac{1}{m_h^4} \Big[ \Big| m_c^2 ( V_{L21}^{ue*} V_{L22}^{ue}
+ 9 V_{L21}^{ue*} V_{R22}^{ue} V_{L22}^{ue*} V_{R22}^{ue} ) \nonumber\\
&\ &\qquad\quad + m_t^2 ( V_{L31}^{ue*} V_{L32}^{ue} + 9 
\frac{m_{\tau}}{m_{\mu}} V_{L31}^{ue*} V_{R33}^{ue} V_{L33}^{ue*}
V_{R32}^{ue} ) \Big|^2 + \Big| L \leftrightarrow R \Big|^2 \Big]
\end{eqnarray}

Experimentally, $B(\mu \rightarrow e \gamma) < 4.9 \times 10^{-11}$ \cite{PDG}.
Parameterising as per Eq.~(\ref{Vhier}), the branching ratio has been plotted
in Fig.~(\ref{ChBEG})
as a function of $\epsilon$ for various values of the Higgs boson mass for the
case in which the charged Higgs boson contribution is dominant.
The corresponding plots for the other cases are similar.
In all these plots the experimental limit is only a couple of orders of 
magnitude greater than the branching ratios corresponding to that region of
parameter space of greatest interest. As for current limits, an assumed value
of $\epsilon = 0.22$ gives the bound $m_H > 92$ GeV on the mass of the
charged Higgs boson.

Again, these calculations can be performed for the tauon as well as the muon.
Once more using the $\epsilon$ parametrisation, we obtain the branching ratios
\begin{eqnarray}
B(\tau \rightarrow \mu \gamma) & \sim & B_{\tau} \frac{\alpha}{96 \pi}
m_t^4 \epsilon^4 \Big[ \Big( \frac{m_{\mu}}{m_{\tau}} \frac{1}{m_H^2}
- \frac{1}{m_+^2} - \frac{m_{\mu}}{m_{\tau}} \frac{1}{m_+^2}
+ 9 \frac{1}{m_-^2} \Big)^2 \nonumber\\
&\ &\qquad\quad + \Big( \frac{1}{m_H^2} - \frac{1}{m_+^2} 
- \frac{m_{\mu}}{m_{\tau}} \frac{1}{m_+^2} + 9 \frac{1}{m_-^2} \Big)^2 \Big] \\
B(\tau \rightarrow e \gamma) & \sim & B_{\tau} \frac{\alpha}{96 \pi}
m_t^4 \epsilon^6 \Big[ \Big( \frac{1}{m_+^2} - \frac{9}{m_-^2} \Big)^2
+ \Big( \frac{1}{m_H^2} - \frac{1}{m_+^2} + \frac{9}{m_-^2} \Big)^2 \Big]
\end{eqnarray}
These branching ratios are $\sim 10^{-8}$ and $\sim 10^{-10}$ respectively
(with masses $\sim$ 100 GeV and $\epsilon \sim 0.1$), compared to the
experimental upper bounds $5.5 \times 10^{-4}$ and $2.0 \times 10^{-4}$.

\subsection{Muon Anomalous Magnetic Moment}
The calculation of the anomalous magnetic moment of the muon to one-loop is
very similar to the above calculation of $\mu \rightarrow e \gamma$. The
total contribution of the various Higgs particles is
\begin{eqnarray}
\Delta a_{\mu} & \approx & \frac{G_F m_{\mu}^2}{24 \sqrt{2} \pi}
\sum_k \Big[ y_{d_k} | V_{Lk2}^{de}|^2 + y_{u_k} |V_{Lk2}^{ue}|^2
+ w_{u_k}^+ \Big( |V_{Lk2}^{ue}|^2 + |V_{Rk2}^{ue}|^2 \Big) \nonumber\\
& & - 9 \sum_{i,l} \frac{m_{e_i}}{m_{\mu}} \sqrt{w_{u_k}^- w_{u_l}^-}
{\rm Re}(V_{Lk2}^{ue*} V_{Rki}^{ue} V_{Lli}^{ue*} V_{Rl2}^{ue}) \Big]
\label{MAG}
\end{eqnarray}

If typical values of the masses and $\epsilon$ are substituted into 
Eq.~(\ref{MAG}), the anomalous magnetic moment is found to be less than
$\sim 10^{-11}$. Experimentally it is known \cite{PDG} that
$-13 \times 10^{-9} < \Delta a_{\mu} < 21 \times 10^{-9}$, so this constraint
is easily satisfied.

\subsection{$b \rightarrow s \gamma$}
The nonleptonic decay $b \rightarrow s \gamma$ is currently of interest
due to the recent detection of this process by the CLEO collaboration
\cite{CLEO}, and is as yet not in disagreement with Standard Model predictions.
The 2HDM contribution has previously been the focus of much investigation
\cite{bsgamma},
and so it is worthwhile considering how the quark-lepton symmetric two Higgs
doublet model differs in its effect on this process.

Using the notation of Grinstein, Springer and Wise \cite{GSW}, the form factors
$C_7 (m_W)$ and $C_8 (m_W)$, referring to the operators for the photonic and
gluonic vertices respectively, at the mass-scale $M_W$, need to be calculated.

In the Standard Model, these form factors are given by
\begin{eqnarray}
C_7 (m_W) & = & - \frac{1}{2} A(x_t) \\
C_8 (m_W) & = & - \frac{1}{2} D(x_t)
\end{eqnarray}
where $x_f \equiv m_f^2 / m_W^2$, and
\begin{eqnarray}
A(x) & \equiv & x \Big[ \frac{\frac{2}{3} x^2 + \frac{5}{12} x - \frac{7}{12}}
{(x-1)^3} - \frac{(\frac{3}{2} x^2 - x)\ln x}{(x-1)^4} \Big] \\
D(x) & \equiv & \frac{x}{2} \Big[ \frac{\frac{1}{2} x^2 - \frac{5}{2} x - 1}
{(x-1)^3} + \frac{3 x \ln x}{(x-1)^4} \Big]
\end{eqnarray}

In this quark-lepton symmetric two-Higgs doublet model,
the dominant contribution comes from the
heavier third generation, in both the quark and the lepton sectors.
Performing the necessary one-loop calculation, again similar to that
for $\mu \rightarrow e \gamma$, the result is (for simplicity only
considering charged Higgs)
\begin{eqnarray}
C_7 (m_W) & = & - \frac{1}{2} A(x_t) - \frac{1}{6} \frac{y_{\tau}}{y_t}
\frac{V_{L23}^{de} V_{L33}^{de*}}{V_{L32}^{ud*}} A(y_t) \\
C_8 (m_W) & = & - \frac{1}{2} D(x_t) - \frac{1}{6} \frac{y_{\tau}}{y_t}
\frac{V_{L23}^{de} V_{L33}^{de*}}{V_{L32}^{ud*}} D(y_t)
\end{eqnarray}
Because $m_W = $ 83 GeV, $m_H >$ 45.3 GeV, and $A(x)$ increases with $x$
(ie. decreases with $m_W$ or $m_H$), $A(y_t) < 3.36 A(x_t)$.
If the mixing matrices are similar to the CKM matrix then
$V_{L23}^{de} V_{L33}^{de*}/V_{L32}^{ud*} \sim 1$ (at any rate it
cannot be too much greater than 1). The deciding factor is then
$y_{\tau}/y_t = m_{\tau}^2/m_t^2 \ll 1$, so the charged Higgs
contribution should be negligible to that from the Standard Model. 
As stated above, this is in agreement with current experimental data.

\section{Conclusion}
If the two Higgs doublet version of the quark-lepton symmetric model 
is correct, then as we have seen, its contribution to various 
leptonic decays should be measurable for a wide range of parameters, 
with only a small improvement in detector precision. As however 
the Higgs boson masses and the mixing between lepton families are unknown, 
it is possible that extreme values (i.e. very diagonal mixing matrices,
or very heavy scalars), could lead to extremely small branching ratios.
Measurement of the above decays is at least one way to obtain
information about these parameters.

\acknowledgments
\section*{}
CCL was supported by an Australian Postgraduate Research Award.
RRV was supported by the Australian Research Council and the University
of Melbourne.

\appendix
\section*{}
The total decay rate for the process $\mu \rightarrow e \gamma$ to one loop,
including both Standard Model and charged and neutral Higgs contributions,
is (assuming small lepton masses)
\begin{eqnarray}
\Gamma (\mu \rightarrow e \gamma) & \approx &
\frac{\alpha G_F^2 m_{\mu}^2}{128 \pi^2} \Big( 1 - \frac{m_e^2}{m_{\mu}^2}
\Big)^3
\Big[ \Big| \sum_{i,k,l} \Big( m_{\mu} F_2 V_{Li2}^{e\nu} V_{Li1}^{e\nu*}
+ m_{\mu} I_A \sqrt{y_{d_k} y_{d_l}} V_{Lk1}^{de*} V_{Rki}^{d\nu}
V_{Rli}^{d\nu*} V_{Ll2}^{de} \nonumber\\
& & + m_e I_B \sqrt{y_{u_k} y_{u_l}} V_{Rk1}^{ue*} V_{Lki}^{u\nu}
V_{Lli}^{u\nu*} V_{Rl2}^{ue} + m_{\nu_i} I_i \sqrt{y_{d_k} y_{u_l}}
V_{Lk1}^{de*} V_{Rki}^{d\nu} V_{Lli}^{u\nu*} V_{Rl2}^{ue} \nonumber\\
& & - \frac{1}{2} m_{\mu} I'_A \sqrt{w_{u_k}^+ w_{u_l}^+}
V_{Lk1}^{ue*} V_{Rki}^{ue} V_{Rli}^{ue*} V_{Ll2}^{ue}
- \frac{1}{2} m_e I'_B \sqrt{w_{u_k}^+ w_{u_l}^+} V_{Rk1}^{ue*} V_{Lki}^{ue}
V_{Lli}^{ue*} V_{Rl2}^{ue} \nonumber\\
& & - \frac{1}{2} m_{e_i} I'_i
\sqrt{w_{u_k}^- w_{u_l}^-} V_{Lk1}^{ue*} V_{Rki}^{ue} V_{Lli}^{ue*}
V_{Rl2}^{ue} \Big) \Big|^2 \nonumber\\
& & + \Big| \sum_{i,k,l} \Big( m_e F_2 V_{Li2}^{e\nu} V_{Li1}^{e\nu*}
+ m_{\mu} I_A \sqrt{y_{u_k} y_{u_l}} V_{Rk1}^{ue*} V_{Lki}^{u\nu}
V_{Lli}^{u\nu*} V_{Rl2}^{ue} \nonumber\\
& & + m_e I_B \sqrt{y_{d_k} y_{d_l}} V_{Lk1}^{de*} V_{Rki}^{d\nu}
V_{Rli}^{d\nu*} V_{Ll2}^{de} + m_{\nu_i} I_i \sqrt{y_{u_k} y_{d_l}}
V_{Rk1}^{ue*} V_{Lki}^{u\nu} V_{Rli}^{d\nu*} V_{Ll2}^{de} \nonumber\\
& & - \frac{1}{2} m_{\mu} I'_A \sqrt{w_{u_k}^+ w_{u_l}^+}
V_{Rk1}^{ue*} V_{Lki}^{ue} V_{Lli}^{ue*} V_{Rl2}^{ue}
- \frac{1}{2} m_e I'_B \sqrt{w_{u_k}^+ w_{u_l}^+} V_{Lk1}^{ue*} V_{Rki}^{ue}
V_{Rli}^{ue*} V_{Ll2}^{ue} \nonumber\\
& & - \frac{1}{2} m_{e_i} I'_i
\sqrt{w_{u_k}^- w_{u_l}^-} V_{Rk1}^{ue*} V_{Lki}^{ue} V_{Rli}^{ue*}
V_{Ll2}^{ue} \Big) \Big|^2 \Big]
\label{BEG0}
\end{eqnarray}
where $F_2$ is the Standard Model contribution \cite{ILim},
\begin{equation}
F_2 \approx\ -\frac{x_{\nu_i}}{4}
\end{equation}
with $x_f \equiv m_f^2/m_W^2$. The other terms in Eq.~(\ref{BEG0}) are
\begin{eqnarray}
I_A & = & \int_0^1 du \int_0^{1-u} dv \, (1-u-v)v/D \approx \frac{1}{12},
\;\; \mbox{for small lepton masses},\nonumber\\
I_B & = & \int_0^1 du \int_0^{1-u} dv \, (1-u-v)u/D \approx \frac{1}{12},
\nonumber\\
I_i & = & \int_0^1 du \int_0^{1-u} dv \, (1-u-v)/D \approx \frac{1}{2},
\nonumber\\
I'_A & = & \int_0^1 du \int_0^{1-u} dv \, (1-u-v)v/D' \approx \frac{1}{6},
\nonumber\\
I'_B & = & \int_0^1 du \int_0^{1-u} dv \, (1-u-v)u/D' \approx \frac{1}{6},
\nonumber\\
I'_i & = & \int_0^1 du \int_0^{1-u} dv \, (u+v)/D' \approx -\frac{3}{2},
\nonumber\\
D & = & (1-u-v)(y_{\nu_i}-1-v y_{\mu}-u y_e) + 1,\nonumber\\
D' & = & (1-u-v)(1-w_{e_i}-v w_{\mu}-u w_e) + w_{e_i}.\nonumber
\end{eqnarray}


\begin{figure}
\caption{Plot of the branching ratio for the process $\mu \rightarrow
e \bar{e} e$ against the mixing matrix parameter $\epsilon$, for a range
of masses of the neutral Higgs particles. The upper line in each pair
corresponds to the mass $m_{\phi}$, and the lower line to $m_{\eta}$
and $m_h$. The experimentally obtained upper
limit on the branching ratio is also indicated.}
\label{NeuBEEE}
\end{figure}

\begin{figure}
\caption{Plot of the branching ratio $B(\mu \rightarrow e \gamma)$ against
$\epsilon$ for various values of the mass of the charged Higgs boson $H^+$,
with the current experimental upper bound 
indicated.}
\label{ChBEG}
\end{figure}


\begin{references}
\bibitem{FootLew} R. Foot and H. Lew, Phys. Rev. {\bf D41}, 3052 (1990);
R. Foot, H. Lew and R. R. Volkas, {\it ibid.} {\bf D44}, 1531 (1991);
{\it ibid.} {\bf D44}, 859 (1991); Int. J. Mod. Phys. {\bf A8}, 983 (1993);
Mod. Phys. Lett. {\bf A8}, 1859 (1993); R. Foot and H. Lew, Nuovo Cim.
{\bf A104}, 167 (1991); Phys. Rev. {\bf D42}, 945 (1990); Mod. Phys. Lett.
{\bf A5}, 1345 (1990); {\it ibid.} {\bf A7}, 301 (1992); G. C. Joshi and
R. R. Volkas, Phys. Rev. {\bf D45}, 1711 (1992); H. Lew and R. R. Volkas,
{\it ibid.} {\bf D47}, 1356 (1993); R. R. Volkas, University of Melbourne
Preprint UM-P-94/19.
\bibitem{LeVolk} Y. Levin and R. R. Volkas, Phys. Rev. {\bf D48}, 5342 (1993).
\bibitem{WeinGlash} S. Glashow and S. Weinberg, Phys. Rev. {\bf D15}, 1958 
(1977).
\bibitem{Hunter} H. E. Haber, G. L. Kane and T. Sterling, Nucl. Phys.
{\bf B161}, 493 (1979); N. G. Deshpande and E. Ma, Phys. Rev. {\bf D18}, 2574
(1978); H. Georgi, Hadronic J. {\bf 1}, 155 (1978); J. F. Donoghue and L.-F. Li,
Phys. Rev. {\bf D19}, 945 (1979); L. F. Abbott, P. Sikivie and M. B. Wise,
Phys. Rev. {\bf D21}, 1393 (1980); B. McWilliams and L.-F. Li, Nucl. Phys.
{\bf B179}, 62 (1981); J. F. Gunion and H. E. Haber, Nucl. Phys. {\bf B272}, 1
(1986). For a review see
J. F. Gunion, H. E. Haber, G. L. Kane and S. Dawson,
{\it The Higgs Hunter's Guide} (Addison-Wesley Publishing Company,
Reading, MA 1990).
\bibitem{ALEPH} ALEPH Collaboration, Phys. Rep. {\bf 216}, 253 (1992).
\bibitem{PDG} Particle Data Group, Phys. Rev. {\bf D45}, S1 (1992).
\bibitem{ILim} T. Inami and C. S. Lim, Prog. Theor. Phys. {\bf 65},
297 (1981); E. Ma and A. Pramudita, Phys. Rev. {\bf D24}, 1410 (1981);
W. J. Marciano and A. I. Sanda, Phys. Lett. {\bf 67B}, 303 (1977).
\bibitem{bsgamma} T. D. Nguyen and G. C. Joshi, Phys. Rev. {\bf D37}, 3220
(1988); T. G. Rizzo, Phys. Rev. {\bf D38}, 820 (1988); R. G. Ellis, G. C. Joshi
and M. Matsuda, Phys. Lett. {\bf B179}, 119 (1986); W.-S. Hou and R. S. Willey,
Phys. Let. {\bf B202}, 591 (1988); V. Barger, M. S. Berger and R. J. N.
Phillips, Phys. Rev. Lett. {\bf 70}, 1368 (1993).
\bibitem{CLEO} CLEO Collaboration, Phys. Rev. Lett. {\bf 71}, 674 (1993).
\bibitem{GSW} B. Grinstein, R. Springer and M. B. Wise, Nucl. Phys. 
{\bf B339}, 269 (1989).
\end{references}
\end{document}